\documentclass[11pt]{article}
\usepackage[T1]{fontenc}
\usepackage[utf8]{inputenc}
\usepackage{tgbonum}
{\Roman{section}}
\usepackage{amsmath,amssymb,slashed,graphicx,xcolor,bbm}
\usepackage[leftcaption]{sidecap}
\numberwithin{equation}{section}
\DeclareMathAlphabet{\boldmathe}{T1}{cmr}{bx}{it}
\newcommand{\mbf}[1]{\boldmathe{#1}}

\newcommand{\vx}{\mbf{x}}

\newcommand{\omegaud}[2]{{\omega^{#1}}_{#2}}
\newcommand{\omegadu}[2]{{\omega_{#1}}^{#2}}
\newcommand{\Omegaud}[2]{{\Omega^{#1}}_{#2}}
\newcommand{\Phiud}[2]{\Phi^{#1}_{\;\;#2}}
\newcommand{\al}{\alpha}
\newcommand{\cL}{\mathcal{L}}
\newcommand{\D}{\mathrm{d}}

\newcommand{\ft}[2]{{\textstyle\frac{#1}{#2}}}
\newcommand{\ftez}{{\textstyle\frac{1}{2}}}
\begin{document}
	\begin{titlepage}
		\title {
		\hfill{\small preprint SIN-PR-80-008}\\[15mm]	
			Post-Newtonian Generation of Gravitational Waves
			in a Theory of Gravity with Torsion}
		\author{
			M. Schweizer, N. Straumann, and A. Wipf
			\footnote{present address:
								Theoretisch-Physikalisches-Institut, Friedrich-Schiller-Universit{\"a}t Jena, 07743 Jena,
								Germany; email:
								wipf@tpi.uni-jena.de}\\
			\emph{\small Institute for Theoretical Physics of the University of Z{\"u}rich,}\\
						\emph{\small Sch{\"o}nberggasse 9, 8001 Z{\"u}rich, Switzerland}
		}
		\maketitle
		\begin{abstract}
			\noindent
		We adapt the post-Newtonian gravitational-radiation methods developed within general
		relativity by Epstein and Wagoner to the gravitation theory with torsion, recently proposed
		by Hehl et al., and show that the two theories predict in this approximation the same
		gravitational radiation losses. Since they agree also on the first post-Newtonian level, they are at the present time -- observationally -- indistinguishable.
		\end{abstract}
		\vskip25mm
		\begin{center}\small
			Published version: General Relativity and Gravitation, Vol. 12 (1980) 951-961\\
			doi: 10.1007/BF00757366\\
			arXiv-ed February 1980 ; \LaTeX-ed May 1, 2023			
		\end{center}
		Uploaded to arXiv to help researchers with limited library facilities
	\end{titlepage}
	\tableofcontents


\section{Introduction}\label{sec:1}
The binary pulsar data obtained by Taylor and coworkers (Taylor \cite{1})
have already ruled out many alternative gravitation theories. Will \cite{2} has
adopted the post-Newtonian gravitational-radiation methods developed within
general relativity by Epstein and Wagoner \cite{3} and by Wagoner and Will \cite{4} to
alternative metric theories of gravitation and thus has shown that in most (if not
all) of these dipole gravitational radiation can also exist. The dipole radiation of
a two-body system, say, is generated by the varying dipole moment of the 
gravitational binding energy and is typically much too large for the binary pulsar
PSR 1913+ 16. There are examples (e.g. the Rosen bimetric theory) where the
dipole gravitational radiation causes the system even to gain energy at a 
relatively high rate and strong dissipative mechanisms would have to be invented to
account for the observed decrease.
In an earlier paper \cite{5} we have shown that the "Poincar\'e gauge theory of
gravity" proposed by Hehl et al. \cite{6} agrees with general relativity on the first
post-Newtonian level. The purpose of the present work is to demonstrate that
the "dipole catastrophe" mentioned above does not occur in this theory and
that -- within the approximation scheme of \cite{3} -- the gravitational radiation loss
is the same as in general relativity. From the observational point of view it is,
therefore, practically impossible to distinguish the theory of Hehi et al. from
general relativity.
In Section \ref{sec:2}, the theory of Hehl et al. (slightly generalized) is briefly 
summarized in the language of differential forms. In Section \ref{sec:3}, we recast the field
equations into a form from which one can deduce easily that the first 
post-Newtonian approximation of the theory agrees with that of general relativity.
In Section \ref{sec:4}, we determine the energy-momentum forms of the gravitational
field, and finally we demonstrate in Section \ref{sec:5} that the post-Newtonian 
generation of gravitational radiation agrees also with general relativity. The results
are summarized in Section \ref{sec:6}

\section{The Poincar\'e Gauge Theory of Hehl et al.}\label{sec:2}
In this section, we give a short description of the gravitation theory which
was proposed by Hehl et al. \cite{6}. In contrast to Reference \cite{6} we use Cartan's
calculus of differential forms and follow the conventions of Trautman \cite{7}. (See
also \cite{5,8,9} .) Let $\theta^\al$ denote an orthonormal tetrad field of 1-forms, and 
$\omegaud{\al}{\beta}$
the connection forms of a metric connection with torsion. The exterior 
covariant derivatives of $\theta^\al$ and $\omegaud{\al}{\beta}$ 
are the torsion forms $\Theta^\al$ and the curvature forms
$\Omegaud{\al}{\beta}(c=1,8\pi G = 1)$. The 1-forms 
$(\theta^\al;\omegaud{\al}{\beta})$ can be considered as gauge 
potentials of the Poincar\'e group because they determine a connection in the 
Poincar\'e bundle (which is a subbundle of the affine bundle). In this interpretation, 
the 2-forms $(\Theta^\al,\Omegaud{\al}{\beta})$ are the corresponding field strengths. 
A slight generalization
of the gravitational Lagrangian which was chosen in \cite{6} reads as follows
\begin{align}
\cL&=\cL_\mathrm{transl}+\cL_\mathrm{rot}\,,\label{2.1}\\
\cL_\mathrm{transl}&=-\frac{1}{2l^2}\Big[
(\Theta^\al\wedge\theta^\beta)\wedge{^*}(\Theta_\beta\wedge\theta_\al)-
\frac{\lambda}{2}(\Theta^\al\wedge\theta_\al)\wedge{^*}(\Theta^\beta\wedge\theta_\beta)
\Big]\,,\label{2.2}\\
\cL_\mathrm{rot}&=-\frac{1}{2k}\Omega^{\al\beta}\wedge{^*}\Omega_{\al\beta}\,.\label{2.3}
\end{align}
Here $l$ is the Planck length and $k$ is a dimensionless coupling constant. Hehl et al.
use $\lambda=0$ in \eqref{2.2}. To the Lagrangian \eqref{2.1} we have to add a matter term,
$L_m$.
Independent variations of $L_g + L_m$ with respect to $\theta^\al$ and 
$\omegaud{\al}{\beta}$ give the field equations. The variation of $L_m$ with respect 
to $\omegaud{\al}{\beta}$ determines the spin density
which is for macroscopic matter very small compared to the energy-momentum
density. For this reason Hehl et al. choose for astronomical systems the 
"translational gauge limit" in which the curvature vanishes
\begin{equation}
	\Omegaud{\al}{\beta}=0\,.\label{2.4}
\end{equation}
Consequently the rotational (Yang-Mills type) part \eqref{2.3} in the Lagrangian (2.1)
has to be dropped. The resulting theory has an obvious geometric interpretation
in a Weitzenb{\"o}ck space. Relative to a teleparallel tetrad the connection forms
$\omegaud{\al}{\beta}$ also vanish and the exterior covariant derivative reduces the ordinary 
exterior derivative. The torsion $\Theta^\alpha$ is then equal to $\D\theta^\alpha$. 
Thus, we use the following total Lagrangian relative to an arbitrary (orthonormal) frame:
\begin{equation}
	\cL=-\ftez(\Theta^\al\wedge\theta^\beta)\wedge{^*}(\Theta_\beta\wedge\theta_\alpha)
	+\ft{\lambda}{4}(\Theta^\al\wedge\theta_\al)\wedge{^*}(\Theta^\beta\wedge\theta_\beta)+
	\cL_m\,,\label{2.5}
\end{equation}
which reduces to
\begin{equation}
	\cL=-\ftez(\D \theta^\al\wedge\theta^\beta)\wedge{^*}(\D\theta_\beta\wedge\theta_\alpha)
	+\ft{\lambda}{4}(\D\theta^\al\wedge\theta_\al)\wedge{^*}(\D\theta^\beta\wedge\theta_\beta)+
	\cL_m\label{2.6}
\end{equation}
for a teleparallel frame.

Now we note that the Hilbert-Einstein Lagrangian can be written in the
following way (up to an exact differential):
\begin{equation}
	\cL=-\ftez(\D \theta^\al\wedge\theta^\beta)\wedge{^*}(\D\theta_\beta\wedge\theta_\alpha)
+\ft{1}{4}(\D\theta^\al\wedge\theta_\al)\wedge{^*}(\D\theta^\beta\wedge\theta_\beta)+
\cL_m\,.\label{2.7}
\end{equation}
This can easily be shown by putting $\D \theta^\al=
\ft{1}{2}C^\al_{\beta\gamma}\theta^\beta\wedge\theta^\gamma$ and expressing both
sides of \eqref{2.7} in terms of $C^\al_{\beta\gamma}$. Thus the gravitational part of 
\eqref{2.6} reduces to the Hilbert-Einstein Lagrangian for $\lambda=1$. 
This remark is crucial for all that follows and was also used in \cite{5}. 
It is not surprising that the gravitational Lagrangian is invariant under local 
Lorentz transformations
\begin{equation}
	\theta^\al(x)\longrightarrow  \Lambda^{\al}_{\;\,\beta}(x)\theta^\beta(x),\qquad
	\Lambda(x)\in \cL^\uparrow_+\,,\label{2.8}
\end{equation}
for $\lambda=1$ only.

Variation of the tetrad fields leads to the field equations. We introduce the
notations:
\begin{align}
\delta\cL_E&=\delta\theta^\al\wedge \epsilon^E_\alpha\,,  \label{2.9}\\
\delta\cL_m&=\delta\theta^\al\wedge  t_\alpha\,,\label{2.10}\\
\delta(\Delta\cL)&=\delta\theta^\al\wedge \Delta\epsilon_\alpha\,, \label{2.11}
\end{align}
where $(\lambda-1)\Delta\cL$ is the difference between the gravitational part of 
\eqref{2.6} and \eqref{2.7}. The explicit expressions of $\epsilon^E_\alpha$ and 
$\Delta\epsilon_\alpha$ are
\begin{align}
\epsilon^\alpha_E\;&=-\D\big[\theta^\beta\wedge{^*}(\D\theta_\beta\wedge\theta^\al) \big]
-\D\theta^\beta\wedge{^*}(\D\theta^\al\wedge\theta_\beta)
+\ftez\D\big\{\theta^\al\wedge{^*}(\D\theta^\beta\wedge\theta_\beta)\big\}\nonumber\\
&\hskip75mm+\ftez\D\theta^\alpha\wedge{^*}(\D\theta^\beta\wedge\theta_\beta)\,,\label{2.12}
\\
\Delta\epsilon^\al&=\D\theta^\al\wedge{^*}(\D\theta^\beta\wedge\theta_\beta)
-\ftez\theta^\alpha\wedge d\, {^*}(\D\theta^\beta\wedge\theta_\beta)\,.\label{2.13}
\end{align}
and the field equations read as follows:
\begin{equation}
	\epsilon^\alpha_E+(\lambda-1)\Delta\epsilon^\alpha=-t^\alpha\,.
	\label{2.14}
\end{equation}
$t^\al$ are the energy-momentum 3-forms of matter. The components of the 
energy-momentum tensor, $T^{\al\beta}$, relative to $\theta^\al$ are given by
\begin{equation}
	{^*}t^\al=T_\beta^{\;\,\al}\theta^\beta\,.\label{2.15}
\end{equation}
The field equations \eqref{2.14} and the Einstein equations have a large family of
common solutions. For any metric of the diagonal form
\begin{equation*}
\D s^2=a_0^2 (\D x^0)^2-a_1^2 (\D x^1)^2-a_2^2 (\D x^2)^2-a_3^2 (\D x^3)^2
\end{equation*}	
with arbitrary functions $a_i$ the forms $\Delta\epsilon^\al$ vanish for the orthonormal 
basis
\begin{equation*}
\theta^0=a_0 \D x^0,\quad\theta^1=a_1 \D x^1,\quad\theta^2=a_2 \D x^2,\quad\theta^3=a_3 \D x^3
\end{equation*}
because $\D\theta^\al\wedge\theta_\al=0$. This proves in particular that all 
spherically symmetric
solutions of the Einstein equations are also solutions of \eqref{2.14} and thus a large
body of astrophysical applications remains unchanged. The Kerr solution is,
however, no vacuum solution of \eqref{2.14} for $\lambda\neq 1$. It is an open problem to
generalize the stationary black hole solution to this case.
\section{The First Post-Newtonian Approximation}\label{sec:3}
For all the further discussions, it will turn out to be useful to split the field
equations \eqref{2.14} into symmetrized and antisymmetrized parts. In contrast to
$\epsilon^\al_E$ and $t^\al$ the forms $\Delta\epsilon^\al$ are not symmetric:
\begin{equation*}
	\Delta\epsilon^\al\wedge\theta^\beta\neq \Delta\epsilon^\beta\wedge\theta^\al
\end{equation*}
because $\Delta\cL$ is not locally Lorentz invariant. The symmetric and antisymmetric
parts of $\Delta\epsilon^\al$ are
\begin{align}
	\Delta\epsilon^\al_s&=-\ftez\,{^*}\big[(\D\theta^\al\wedge\theta^\beta+\D\theta^\beta\wedge \theta^\al)\wedge {^*}(\D\theta^\gamma\wedge\theta_\gamma)\big]\eta_\beta\,,\label{3.1}
	\\
	\Delta\epsilon^\al_a&=-\ftez\,{^*}\big\{
	\D[\theta^\al\wedge\theta^\beta\wedge {^*}(\D\theta^\gamma\wedge\theta_\gamma)]
	\big\}\eta_\beta\,,\label{3.2}
\end{align}
where $\eta^\beta={^*}\theta^\beta$. Hence the field equations 
\eqref{2.14} are equivalent to the pair of equations
\begin{align}
\epsilon^\alpha_E-\ftez (\lambda-1)&\,{^*}\big[
(\D\theta^\al\wedge\theta^\beta)+\D\theta^\beta\wedge\theta^\al)\wedge {^*}
(\D\theta^\gamma\wedge\theta_\gamma)\big]	\eta_\beta=-t^\al\,,\label{3.3}\\
&(\lambda-1)\D\big[
\theta^\al\wedge\theta^\beta\wedge{^*}(\D\theta^\gamma\wedge\theta_\gamma)
\big]=0\,.\label{3.4}
\end{align}
Now we expand the teleparallel frames $\theta^\al$ in terms of a coordinate basis
\begin{equation}
	\theta^\al=\D x^\alpha+\Phiud{\al}{\beta}\D x^\beta\label{3.5}
\end{equation}
and decompose $\Phi_{\al\beta}$ into its symmetric and antisymmetric pieces
\begin{align}
&\Phi_{\al\beta}=\phi_{\al\beta}+a_{\al\beta}\,,\label{3.6}\\
\phi_{\al\beta}&=\Phi_{(\al\beta)},\qquad a_{\al\beta}=\Phi_{[\al\beta]}\,.\label{3.7}
\end{align}
Let us first consider the linearized approximations of \eqref{3.3} and \eqref{3.4}. 
The second term on the left-hand side of \eqref{3.3} contains obviously no linear terms. 
Furthermore, the antisymmetric field $a_{\al\beta}$ drops out identically in the linearized part
of $\epsilon^\alpha_E$ (as a consequence of the local Lorentz invariance of \eqref{2.7}). 
Thus equation \eqref{3.3} is reduced to the linearized Einstein equation for 
$\phi_{\alpha\beta}$. For $\lambda=1$ (i.e., general relativity) equation \eqref{3.4} 
is empty; $a_{\alpha\beta}$ is just a gauge degree of freedom.
For $\lambda\neq 1$ equation \eqref{3.4} leads in the linearized approximation to 
the decoupled source free equation
\begin{equation}
\Box\, a^{\al\beta}+a^{\lambda\al,\beta}_{\hskip6mm\lambda}-
a^{\lambda\beta,\al}_{\hskip6mm \lambda}=0\label{3.8}
\end{equation}
for $a^{\al\beta}$ which is invariant under the gauge transformation
\begin{equation}
	a_{\al\beta}\longrightarrow a_{\al\beta}+\xi_{\al,\beta}-\xi_{\beta,\al}\,.
	\label{3.9}
\end{equation}
Imposing the gauge condition ${a^{\al\beta}}_{,\beta}=0$, 
we are left with $\Box\,a^{\al\beta}=0$. In the 
Newtonian approximation $a_{\al\beta}$ vanishes in this gauge and the theory gives the correct
Newtonian limit. Next we show that the first post-Newtonian approximation of
\eqref{3.3} and \eqref{3.4} agrees with that of general relativity. (This was shown already 
in \cite{5}, but the following discussion is simpler.) Since the term proportional to 
$(\lambda-1)$ in \eqref{3.3} is of higher order in $\Phi_{\al\beta}$, 
it contains in the first post-Newtonian approximation at most 
quadratic expressions of the Newtonian approximation of
$\phi_{\al\beta}$. But these vanish identically because $\D\theta^\al\wedge\theta_\al$ 
is already quadratic in $\phi_{\al\beta}$.
(The linear term vanishes identically.) Hence \eqref{3.3} reduces to the first 
post-Newtonian approximation of general relativity. For the discussion of equation
\eqref{3.4}, we note first that the Newtonian approximation, $\phi^{\text{(N)}}_{\al\beta}$,
of $\phi_{\al\beta}$ is diagonal
in a suitable coordinate system: $\phi^{\text{(N)}}_{\al\beta}=\delta_{\al\beta}\Phi$, $\Phi$ = Newtonian potential. From this, one concludes easily that the quadratic terms 
in $\phi^{\text{(N)}}_{\al\beta}$ in \eqref{3.4} vanishes identically
and thus the first post-Newtonian approximation of \eqref{3.4} reduces to 
equation \eqref{3.8} for the post-Newtonian approximation, $a^\text{PN}_{\al\beta}$, 
od $a_{\al\beta}$. This implies that $a^\text{PN}_{\al\beta}$
vanishes also in a suitable gauge and hence our claim is proven. Hehl
and Nitsch \cite{10} have shown that the post-post-Newtonian approximation no
longer agrees with that of general relativity. The deviations are, however, 
unmeasurably small.
\section{Conservation Laws}\label{sec:4}
We start by writing the symmetric field equation (3.3) in arbitrary (not
necessarily teleparallel or orthonormal) frames:
\begin{equation}
	\epsilon^\al_E-\ftez (\lambda-1)\,{^*}\big[
	(\Theta^\al\wedge\theta^\beta+\Theta^\beta\wedge\theta^\al)\wedge {^*}(\Theta^\gamma\wedge
	\theta_\gamma)\big]\eta_\beta=-t^\al\,. \label{4.1}
\end{equation}
For $\epsilon^\al_E$ we use the "Landau decomposition" derived in \cite{11}
\begin{equation}
	\epsilon^\al_E=\big[
	2(-g)^{1/2}\big]^{-1}\cdot \D\big[(-g)^{1/2}
	\stackrel{\text{\tiny LC}}{\omega}\!_{\beta}^{\;\;\gamma}\!
	\wedge\,\eta^{\alpha\beta}_{\hskip4mm\gamma}\big]+t^\al_\mathrm{LL}\,.\label{4.2}
\end{equation}
Here ${\stackrel{\text{\tiny LC}}{\omega}}\,\!_{\beta}^{\;\;\gamma}$
are the Levi-Civith connection forms and $t^\al_\mathrm{LL}$ are the Landau-Lifschitz
energy-momentum forms of the metric field given explicitly by
\begin{equation}
t^\al_\mathrm{LL}=-\ftez \eta^{\al\beta\gamma\delta}
\big(
\stackrel{\text{\tiny LC}}{\omega}_{\sigma\beta}\wedge\,
{\stackrel{\text{\tiny LC}}{\omega}}\,\!^{\sigma}_{\;\;\gamma}\wedge\theta_\delta
-
\stackrel{\text{\tiny LC}}{\omega}_{\beta\gamma}\wedge
\stackrel{\text{\tiny LC}}{\omega}_{\sigma\delta}\wedge\theta^\sigma\big)\,.\label{4.3}
\end{equation}
Relative to a coordinate basis $\theta^\al=\D x^\al$ the 3-forms 
$t^\al_\mathrm{LL}$ are symmetric. Inserting \eqref{4.2} into \eqref{4.1} 
the symmetric field equation takes the form
\begin{equation}
	-\ftez \D\big[
	(-g)^{1/2}\stackrel{\text{\tiny LC}}{\omega}_{\beta\gamma}\wedge
	\eta^{\al\beta\gamma}\big]=(-g)^{1/2}\tau^\al\,,\label{4.4}
\end{equation}
where
\begin{align}
	\tau^\al&=t^\al+t^\al_\mathrm{LL}-(\lambda-1)\Delta t^\al\,,\label{4.5}\\
	\Delta t^\al&=\ftez {^*}\big[
	(\Theta^\al\wedge\theta^\beta+\Theta^\beta\wedge\theta^\al)\wedge
	{^*}(\Theta^\gamma\wedge\theta_\gamma)\big]\eta_\beta\,.\label{4.6}
\end{align}
It may be useful to note that the left-hand side of \eqref{4.4} relative to a coordinate
basis can be expressed in terms of the Landau-Lifschitz superpotential as follows:
\begin{equation*}
	\D\big[
	(-g)^{1/2}\stackrel{\text{\tiny LC}}{\omega}_{\beta\gamma}\wedge\eta^{\mu\beta\gamma}
	\big]=(-g)^{1/2}{H^{\mu\al\nu\beta}}_{,\al}\,\eta_\nu\,,
\end{equation*}
where
\begin{equation*}
	H^{\mu\al\nu\beta}={\hat g}^{\mu\nu}{\hat g}^{\al\beta}-{\hat g}^{\mu\beta}{\hat g}^{\nu\al}
\end{equation*}
with
\begin{equation*}
	{\hat g}^{\mu\nu}=(-g)^{1/2} g^{\mu\nu}\,.
\end{equation*}
As in general relativity, the $t^\al$ are interpreted as the total energy-momentum
forms. By construction, they are symmetric relative to a coordinate basis:
\begin{equation}
\tau^\al\wedge\D x^\beta=\tau^\beta\wedge\D x^\al\,.\label{4.7}
\end{equation}
From \eqref{4.4} we conclude that the field equations imply the conservation laws
\begin{equation}
	\D\big[(-g)^{1/2}\tau^\al\big]=0\,.\label{4.8}
\end{equation}
The last two equations imply
\begin{equation}
	\D\big[(-g)^{1/2} \mathrm{M}^{\al\beta}\big]=0\,,\label{4.9}
\end{equation}
where 
\begin{equation}
	\mathrm{M}^{\al\beta}=x^\al\tau^\beta-x^\beta\tau^\al\label{4.10}
\end{equation}
is the total angular momentum density. For isolated systems the total momentum
\begin{equation}
	\mathrm{P}^\al=\int_\Sigma (-g)^{1/2}\tau^\al\label{4.11}
\end{equation}
and the total angular momentum
\begin{equation}
	\mathrm{J}^{\al\beta}=\int_\Sigma (-g)^{1/2}\mathrm{M}^{\al\beta}\label{4.12}
\end{equation}
($\Sigma$: spacelike surface) can be expressed with the help of the field equations 
\eqref{4.4} in terms of flux integrals at infinity
\begin{align}
\mathrm{P}^\al&=-\ftez \oint (-g)^{1/2}\,\stackrel{\text{\tiny LC}}{\omega}_{\beta\gamma}
\wedge \eta^{\al\beta\gamma}\,,\label{4.13}\\
\mathrm{J}^{\al\beta} &=\ftez \oint (-g)^{1/2}\,
	\big[(x^\al\eta^\beta_{\;\;\sigma\gamma}-x^\beta\eta^\al_{\;\;\sigma\gamma})\wedge
	\stackrel{\text{\tiny LC}}{\omega}\!^{\sigma\gamma}+\eta^{\al\beta}
\big]\,.\label{4.14}
\end{align}
As usual, the coordinates have to be asymptotically Lorentzian. 
$\mathrm{P}^\al$ and $\mathrm{J}^{\al\beta}$
transform like Lorentz tensors under coordinate transformations which preserve
this property. Clearly, the flux integrals \eqref{4.13} and \eqref{4.14} are the same as in
general relativity. Since we have not seen equation \eqref{4.14} in the literature, we
derive it in the Appendix.

\section{Post-Newtonian Generation of Gravitational Radiation}\label{sec:5}

In this section we adapt the method of Epstein and Wagoner \cite{3} and show
that the post-Newtonian generation of gravitational radiation is the same as in
general relativity.

First we have to bring the field equations \eqref{3.3} and \eqref{3.6} into a convenient
form by separating explicitly the linear terms in the fields $\phi_{\al\beta}$ 
and $a_{\al\beta}$.

We have already noted in Section \ref{sec:3} that the term proportional to 
$(\lambda-1)$ in \eqref{3.3} contains no linear terms and that the field 
$a_{\al\beta}$ does not appear in the linearized part of $\epsilon^\al_E$.
Hence we only have to split off the linear part in the Einstein form
\begin{equation}
	\epsilon^\al_E=\epsilon^\al_L+\epsilon^\al_Q\,.\label{5.1}
\end{equation}
The linear part, $\epsilon^\al_L$, is given by
\begin{equation}
	\epsilon^\al_L=-G^{\al\beta}_L \eta_\beta\,, \label{5.2}
\end{equation}
where $G^{\al\beta}_L$ is the linearized Einstein tensor
\begin{equation}
-G_{L\beta}^\al=\Box(\phi^\al_{\;\,\beta}-\delta^\al_{\;\,\beta}\phi^\gamma_{\;\,\gamma})
+\delta^\al_{\;\,\beta}{\phi^{\lambda\sigma}}_{,\lambda\sigma}+
{\phi^{\lambda}_{\;\,\lambda}}^{,\al}_{\;\;\,\beta}
-({\phi^{\al\lambda}}_{,\beta}+{\phi^\lambda_{\;\,\beta}}^{,\al})_{,\lambda}\,.
\label{5.3}
\end{equation}
The quadratic and higher-order terms are easily obtained from \eqref{2.12}:
\begin{subequations}
\begin{equation}
	\epsilon^\al_Q=\Delta \epsilon^\alpha_S+\{A^{\beta\al}+B^{\beta\al}+C^{\beta\al} \}\eta_\beta
	\,,	\label{5.4a}
\end{equation}
where
\begin{align}
A^{\nu\mu}&=2{^*}\big[\theta^\nu\wedge\D\theta^\al\wedge{^*}(\D\theta_\al\wedge\theta^\mu)
\big]\,, \label{5.4b}\\	
B^{\nu\mu}&= \ftez {^*}\big[ \theta^\nu\wedge\D\theta^\al\wedge{^*}
\D(\theta^\mu\wedge\theta_\al)+\theta^\mu\wedge\D\theta^\al\wedge{^*}
\D(\theta^\nu\wedge\theta_\al)\big]\,,
\label{5.4c}\\
C^{\nu\mu}&=-\ftez {^*}\big[
\theta^\nu\wedge\theta^\al\wedge\D{^*}(\D\theta_\al\wedge\Phi^\mu_{\;\,\lambda}\D x^\lambda)
+(\D x^\nu\wedge\Phi^\al_{\;\,\sigma}\D x^\sigma+\Phi^\nu_{\;\,\lambda}\D x^\lambda\wedge 
\D x^\al\nonumber\\
&\hskip5mm +\Phi^\nu_{\;\,\lambda}\Phi^\al_{\;\,\sigma}
\D x^\lambda\wedge\D x^\sigma)\wedge \D{^*}(\D\theta_\al\wedge\D x^\mu)\nonumber\\
&\hskip5mm
+\theta^\mu\wedge\theta^\al\wedge \D{^*}(\D\theta_\al\wedge\Phi^\nu_{\;\,\lambda}
\D x^\lambda)+(\D x^\mu\wedge\Phi^\al_{\;\,\lambda}\D x^\lambda +
\Phi^{\mu}_{\;\,\sigma}\D x^\sigma\wedge\D x^\al\nonumber\\
&\hskip5mm +\Phi^\mu_{\;\,\sigma}\Phi^\al_{\;\,\lambda}\D x^\sigma\wedge\D x^\lambda) 
\wedge \D{^*}(\D\theta_\al\wedge\D x^\nu)
\big]\,.
 \label{5.4d}
\end{align}
\end{subequations}
The decomposition (5.4) is equivalent to equation (31) of Reference \cite{3} and is
quite useful for practical (post-Newtonian) calculations.
Inserting \eqref{5.1} into \eqref{3.3} gives
\begin{equation}
	G^{\al\beta}_L=T^{\al\beta}_\mathrm{eff}\label{5.5}
\end{equation}
with
\begin{equation}
	T^{\al\beta}_\mathrm{eff}\eta_\beta= t^\al+\epsilon^\al_Q+(\lambda-1)\Delta\epsilon^\al_S\,,
	\label{5.6}
\end{equation}
where $\Delta\epsilon^\al_S$ is given by \eqref{3.1} (and contains no linear terms). 
The last term in \eqref{5.6} is absent in general relativity. We also rewrite \eqref{3.4} 
(for $\lambda\neq 1$) in a similar way:
\begin{equation}
\Box\,a^{\al\beta}+{a^{\lambda\al,\beta}}_\lambda-{a^{\lambda\beta,\al}}_\lambda=
A^{\al\beta}_\mathrm{ef}\,,\label{5.7}
\end{equation}
where $A^{\al\beta}_\mathrm{ef}$ collects the quadratic and higher-order terms of
\begin{equation*}
	\ftez \D\big[\theta^\al\wedge\theta^\beta\wedge {^*}(\D\theta^\gamma\wedge\theta_\gamma)
	\big]\,.
\end{equation*}
For outgoing-wave boundary conditions, we obtain from \eqref{5.5} and 
\eqref{5.7} the integral equations
\begin{align}
	\phi_{\al\beta}(t,\vx)&=
	-\frac{1}{4\pi}\int \frac{\hat T_{\al\beta}(t-\vert\vx-\vx'\vert,\vx')}{\vert \vx-\vx'\vert}
	\D^3x'+\xi_{\al,\beta}+\xi_{\beta,\al}\,,
	\label{5.8}\\
	(\hat T^\al_{\;\,\beta}&={T_\mathrm{eff}}^\al_{\;\,\beta}
	-\ftez \delta^\al_{\;\,\beta}
	{T_\mathrm{eff}}^\lambda_{\;\,\lambda})\nonumber\\
	a_{\al\beta}(t,\vx)&=\frac{1}{4\pi}\int \frac{\hat A_{\al\beta}^\mathrm{eff}(t-\vert\vx-\vx'\vert,\vx')}{\vert \vx-\vx'\vert}
	\D^3x'+\eta_{\al,\beta}-\eta_{\beta,\al}\,.
	\label{5.9}
\end{align}
The gauge terms added on the right-hand sides of \eqref{5.8} 
and \eqref{5.9} are, of course, not determined.

Let us now consider the fields $\phi_{\al\beta}$ and $a_{\al\beta}$ 
far from the source ($r=\vert\vx\vert\gg R=$"size" of source,
$\vert\phi_{\al\beta}\vert\ll 1, \vert a_{\al\beta}\vert \ll 1$) 
where the radiation is detected.
In the approximation scheme of Reference \cite{3}, one first expands the 
$1/r$ terms  with respect to the retardation within the source (slow motion). 
In the next step, one constructs a post-Newtonian expansion of the sources 
$T^{\al\beta}_\mathrm{eff}$  and $A^{\al\beta}_\mathrm{eff}$.
The necessary orders in the post-Newtonian expansion parameter which are
needed for the various pieces are described in \cite{3} and will not be repeated. In
this approximation, the effective sources are contained within the near zone.
We may then employ \eqref{5.8} and \eqref{5.9} for field points within this region as well,
again in terms of a post-Newtonian expansion. But the near-zone post-Newtonian
expansion was already studied in Section \ref{sec:3}. To the necessary orders 
$a_{\al\beta}$ vanishes and $\phi_{\al\beta}$ agrees with general relativity if we 
impose suitable gauge conditions.
[$A^{\al\beta}_\mathrm{eff}$ and the last term in \eqref{5.6} 
for $T^{\al\beta}_\mathrm{eff}$ both vanish in the near-zone 
post-Newtonian approximation.] Hence, we conclude from \eqref{5.8} and \eqref{5.9} that
far away from the source $a_{\al\beta}$ is just equal to a gauge term and that
$\phi_{\al\beta}$ agrees also there with general relativity. 
But in this weak field region, the equation for the a field decouples and has a 
separate gauge invariance. 
Thus the a field can be gauged away.

Finally we note that the modified energy-momentum $t^\al_\mathrm{LL}-(\lambda-1)\Delta t^\al$
[see equation \eqref{4.5}] does not change the gravitational energy loss because 
$\Delta t^\al$ is cubic in $\phi_{\al\beta}$. 
This follows from \eqref{4.6} and from the fact that the factor
$\D\theta^\gamma\wedge\theta_\gamma$ in \eqref{4.6} is already quadratic in 
$\phi_{\al\beta}$. (The linear term vanishes identically.) 
Taken all together, this proves our claim at the beginning of this section.

\section{Summary}\label{sec:6}
The gravitational theory with torsion corresponding to the one-parameter
family of Lagrangians \eqref{2.6} has many exact solutions in common with general
relativity. We have shown that they agree also with general relativity on the first
post-Newtonian level (but not in higher orders.). The main new result of this
paper is contained in Section \ref{sec:5}, where we demonstrate that even the 
post-Newtonian generation of gravitational waves (developed within general 
relativity by Epstein and Wagoner \cite{3} ) is the same as in general relativity. In 
particular, the "dipole catastrophe," described in the Introduction, which occurs
in many alternative metric theories of gravitation, is absent. Therefore, the 
results obtained in \cite{3} and \cite{4} also hold in the theory by 
Hehl et al. \cite{6}. At the
present time, this theory can thus not be distinguished observationally from
general relativity.

This is, of course, only true if the approximation scheme of Reference \cite{3}
is numerically reliable. Various authors (see, e.g., \cite{12} and references therein)
have criticized the presently existing approximation methods for treating the
radiation problem. We are aware of the critical questions that have been raised,
but we think that the method used here is physically plausible.

Apart from aesthetic arguments, we see no way to favor the Lagrangian with
$\lambda=1$ in \eqref{2.6}, i.e., general relativity.

\appendix
\renewcommand{\thesection}{A}
\section*{Appendix}\label{sec:7}
In this Appendix, we derive the expression \eqref{4.14} for the angular 
momentum. From the definition \eqref{4.10} and the field equations in the form 
\eqref{4.4} we conclude that
\begin{align}
	(-g)^{1/2} \mathrm{M}^{\sigma\al}&=
	\ftez (x^\sigma \D h^\al-x^\al\D h^\sigma)\nonumber\\
&=\ftez \D (x^\sigma h^\al-x^\al h^\sigma)-\ftez (\D x^\sigma \wedge h^\al
-\D x^\al\wedge h^\sigma)\,,\label{A.1}
\end{align}	
where
\begin{equation}
	h^\al=-(-g)^{1/2}(\omega^{\beta\gamma}\wedge {\eta^\al}_{\beta\gamma})\,.\label{A.2}
\end{equation}
In this Appendix $\omegaud{\al}{\beta}$ denote always the Levi-Civit{\`a} connection forms. Now we
write also the last term in 
\eqref{A.1} as an exact differential. We have
\begin{align*}
\D x^\sigma\wedge h^\al-\D x^\al\wedge h^\sigma&=
(-g)^{1/2}\omega^{\beta\gamma}\wedge\D x^\sigma\wedge
{\eta^\al}_{\beta\gamma}-(\al\longleftrightarrow \sigma)\\
&=(-g)^{1/2}\big(\omegadu{\beta}{\sigma}\wedge\eta^{\al\beta}
+\omegaud{\sigma}{\beta}\wedge\eta^{\beta\al}\\
&\hskip18mm-\omegadu{\beta}{\al}\wedge\eta^{\sigma\beta}
-\omegaud{\al}{\beta}\wedge\eta^{\beta\sigma}\big)\,.
\end{align*}
Here we use
\begin{equation*}
	\D\eta^{\sigma\al}+\omegaud{\sigma}{\beta}\wedge
	\eta^{\beta\al}+\omegaud{\al}{\beta}\wedge\eta^{\sigma\beta}
	=0
\end{equation*}
and obtain
\begin{equation*}
\D x^\sigma\wedge h^\al-\D x^\al\wedge h^\sigma=
(-g)^{1/2}\big[\omegadu{\beta}{\sigma}\wedge \eta^{\al\beta}
-(\sigma\longleftrightarrow \al)-\D\eta^{\sigma\al}\big]\,.
\end{equation*}
But 
\begin{equation*}
\omegadu{\beta}{\sigma}	\wedge \eta^{\al\beta}=
\Gamma_{\beta\mu}^{\;\,\sigma}\,\D x^\mu\wedge \eta^{\al\beta}
={\Gamma_\beta}^{\beta\sigma}\eta^\al
-{\Gamma_\beta}^{\al\sigma}\eta^\beta
\end{equation*}	
and hence 
\begin{equation*}
	\D x^\sigma\wedge h^\al-\D x^\al\wedge h^\sigma=
	(-g)^{1/2}\big(\Gamma^{\beta\sigma}_\beta \eta^\al
	-\Gamma^{\beta\al}_\beta \eta^\sigma-\D \eta^{\sigma\al}\big)\,.
\end{equation*}
If we use the expression
\begin{equation*}
\Gamma^{\beta\sigma}_\beta=	(-g)^{1/2}\cdot g^{\mu\sigma}\partial_\mu (-g)^{1/2}
\end{equation*}
then we find easily
\begin{equation}
\D x^\sigma\wedge h^\al-\D x^\al\wedge h^\sigma=
-\D\big[(-g)^{1/2}\cdot\eta^{\sigma\al}\big]\,.
\label{A.3}
\end{equation}
With this result and \eqref{A.2} equation \eqref{A.1} becomes
\begin{equation}
	(-g)^{1/2}\mathrm{M}^{\sigma\al}=
	\ftez \D\big\{
	(-g)^{1/2}[\eta^{\sigma\al}+(x^\sigma{\eta^\al}_{\beta\gamma}
	-x^\al{\eta^\sigma}_{\beta\gamma})\wedge\omega^{\beta\gamma}]
	\big\}\,.
\label{A.4}
\end{equation}
Inserting this into (4.14) and using Stokes' theorem finally gives equation (4.14)

\section*{Acknowledgement}
We thank Dr. M. Camenzind for many useful discussions


\begin{thebibliography}{10}
\bibitem{1} Taylor, J. H., Hulse, R. A., Fowler, L. A., Gullahorn, G. E., and Rankin, J. M. (1976). Astrophys. J., 206, L53; Taylor, J. H. (1978). Talk given at the Ninth Texas Symposium on Relativistic Astrophysics (M{\"u}nchen, 1978); Taylor, J. H., Fowler, L. A., and McCullough, P. M. (1979). Nature, 277,437.
\bibitem{2} Will, C. M. (1977). Astrophys. J., 214, 826-839; (1978). Talk given at the Ninth Texas Symposium on Relativistic Astrophysics (M{\"u}nchen, 1978).
\bibitem{3} Epstein, R., and Wagoner, R. V. (1975). Astrophys. J., 197,717.
\bibitem{4} Wagoner, R. V., and Will, C. M. (1976). Astrophys. J., 210, 764.
\bibitem{5} Schweizer, M., and Straumann, N. (1979). Phys. Lett., 71A, 493.
\bibitem{6} Hehl, F. W., Ne'eman, Y., Nitsch, J., and Von der Heyde, P. (1978). Phys. Lett., 788, 102; Hehl, F. W., Nitsch, J., and Von der Heyde, P. (1980). Einstein Commemorative Volume, ed. Held, A. Plenum, New York (to appear); Hehl, F. W. (1980). Four
Lectures on Poincar{\'e} Gauge Field Theory, Proceedings of the International School of
Cosmology and Gravitation, Erice, May 1979, eds. Bergmann, P. G. and de Sabbata, V.
Plenum, New York (to appear).
\bibitem{7} Trautman, A. (1973). Symposia Mathematica, Vol. 12, Academic, New York, p. 139.
\bibitem{8} Rumpf, H. (1978). Z. Naturforsch., 33a, 1224.
\bibitem{9} Wipf, A. (1979). Diplomarbeit, "Analyse einer Poincar\'e Eichfeldtheorie der Gravitation und Emission yon Gravitationsstrahlung in derselben," Institute for Theoretical Physics, Sch{\"o}nberggasse 9, University of Z{\"u}rich, CH-8001 Z{\"u}rich.
\bibitem{10} Hehl, F. W., and Nitsch, J. (1979). University of Cologne, preprint; Nitsch, J. (1980).
Seminar Talk, Proceedings of the International School of Cosmology and Gravitation,
Erice, May 1979, eds. Bergmann, P. G., and de Sabbata, V. Plenum, New York (to
appear).
\bibitem{11} Thirring, W. (1978). Lehrbuch der Mathematischen Physik, Vol. 2, Springer-Verlag, Wien.
\bibitem{12} Ehlers, J. (1978). "Isolated Systems in General Relativity," Talk given at the Ninth Texas Symposium on Relativistic Astrophysics, (M{\"u}nchen, 1978).	
\vfill\eject
\end{thebibliography}
\end{document}